# Parcels of Universe or why Schrödinger and Fourier are so relatives?

Marco Frasca, Alfonso Farina *LFellow, IEEE*

This paper is about the surprising connection between the Fourier heat equation and the Schrödinger wave equation. In fact, if the independent "time" variable in the heat equation is replaced by the time variable multiplied by $i = \sqrt{-1}$, the heat equation becomes the Schrödinger equation. Two quite different physical phenomena are put in close connection: the heat diffusion in a material and the probability amplitude of particles in an atom. It is a fact of life that the movements of a small particle floating randomly in a fluid, the well-known Brownian motion, is regulated by the Fourier equation while the probabilistic behavior of the matter around us, the quantum world, is driven by the Schrödinger equation but no known stochastic process seems at work here. The apparent simplicity of the formal connection by a "time-rotation", a Wick rotation as it is commonly known, seems to point otherwise. Why this connection? Is there any physical intuitive explanation? Is there any practical value? In this paper, the authors attempt to shed some light on the above questions. The recent concept of volume quantization in noncommutative geometry, due to Connes, Chamseddine and Mukhanov, points again to stochastic processes also underlying the quantum world making Fourier and Schrödinger strict relatives.

## I. INTRODUCTION

In reference [1], it is shown that the numbers along a row of a Tartaglia-Pascal triangle goes to fit a Gaussian function. In the paper [2], the authors obtain a kind of quantum Tartaglia-Pascal triangle that appears to be the "square root" of the classical one. A square root could entail also imaginary values and it is here where the deep connection between Brownian motion and quantum world starts.

The diffusion heat (Fourier) equation is generally written in the form $\frac{\partial P}{\partial t} = D \frac{\partial^2 P}{\partial x^2}$ where $D$ is the diffusion coefficient and $P$ the probability distribution of the Brownian motion[1]. The Schrodinger equation is written in the form $i\frac{\partial \psi}{\partial t} = -\frac{\hbar^2}{2m}\frac{\partial^2 \psi}{\partial x^2}$ where $\psi$ is the wave function, $\hbar$ the Planck constant divided by $2\pi$ and $m$ the mass of the particle. These equations are formally very similar and can be transformed into each other provided we change the time variable to $t \to it$, what physicists call a Wick rotation [3]. This apparently innocuous modification is indeed a drastic change as now $\psi$ is a complex function while $P$ is a well-behaving real probability distribution. So, a question naturally arises: Where does the imaginary time come from?

The solution of the Fourier equation, describing a probabilistic effect as the scattering of a small particle by the molecules of a fluid[2], has a typical Gaussian form that arises naturally from the asymptotic form of the binomial coefficients, the Tartaglia-Pascal triangle. As shown in [2], a Gaussian probability distribution in a quantum world represents a well-localized particle that, evolving in time, loses such a precise localization, but this Gaussian pdf is the square of the wave function. Therefore, we can connect the binomials of the Fourier equation with the square root of the probability distribution of a moving free particle in a quantum world. However, this entails complex values. This mapping[3] was proved in [2] and is the starting point for a new view on stochastic processes and a parceled world.

So, the rather stunning conclusion is that there exists a square root of the Tartaglia-Pascal triangle, a quantum Tartaglia-Pascal triangle (QTPT), that represents a completely new complex-valued four dimensional figure (x,y,z,t) that lives in the realm of quantum mechanics[4].

Our universe, as we currently understand it, has a peculiar mathematical structure. It is a Riemann manifold with the Hausdorff property [5]. One can always have open sets without intersection with points inside. No pathology whatsoever. A Riemann manifold is deeply connected to stochastic processes. In two dimensions, it can be always reconstructed from Brownian motion [6]. However, there is a deeper connection as shown by Alain Connes, Ali Chamseddine and Viatcheslav Mukhanov. In essence a

---

Marco Frasca is currently working in MBDA Italia S.p.A. at Seeker Division (Rome).

Alfonso Farina was Senior VP Selex-Sistemi Integrati (retired), now consultant of Land&Naval Division of Leonardo Company (Rome).

[1] In Ref.[1] the heat equation was written in the form $u_t(x,t) - \gamma \cdot u_{xx}(x,t) = 0$ where $u$ was equivalent to our $P$ and represented the distribution of the temperature in a body and $\gamma$ was the diffusion constant in the given body. In this paper, $D$ is given by the microscopic motion of atoms colliding with a Brownian particle as devised by Einstein [4].

[2] Note that the effect arises by the averaged squared velocity of the molecules that gives the overall temperature of the fluid itself.

[3] The mapping we proved states: "*There exists a discrete mapping onto the wave function that solves the Schrödinger equation for a free particle via the Tartaglia-Pascal triangle. Such a mapping gives complex-valued probability amplitudes whose squares are the binomial coefficients*."

[4] The correspondence, as given in [2], is between the binomial coefficient $\binom{n}{k}$ and the quantum analog $\sqrt{\binom{n}{k}} 2^{-\frac{n}{2}} e^{i\left(\left(k-\frac{n}{2}\right)^2 \sqrt{\frac{n}{4}-1} - \frac{1}{2}\tan^{-1}\sqrt{\frac{n}{4}-1}\right)}$.

Riemann manifold can be reconstructed by two sets of small volumes [7]-[8]. So, motion between such parcels, by a particle that is able to sense them, can be assimilated to a Brownian motion.

Where do such small volumes come from? The idea comes out from the noncommutative geometry uncovered by Alain Connes in the last century [9]. The idea behind noncommutative geometry can be traced back to the deep connection between algebra and geometry as initially conceived by Descartes. It is well-known that whatever algebraic expression one considers there is a corresponding geometrical object of it. E.g., the equation $x^2 + y^2 = 1$ represents a circle on a Cartesian plane. Nevertheless, all this works fine because complex numbers have the property to commute each other. We learned over the last century that our world does not seem to agree with this property at its foundations. Werner Heisenberg initially conceived this and it is now our understanding of the behavior of elementary particles. In this case, we have what Paul Dirac called q-numbers and momentum $p$ and position $q$ do not commute yielding the famous equation $qp - pq = i\hbar$ (otherwise stated $[q,p] = i\hbar$ where $[,]$ is the commutator) from which the uncertainty principle derives. The reason is that such q-numbers are matrices with an infinite number of elements or, better stated, operators acting on a Hilbert space.

Why such noncommutative behavior changes the behavior of particles? We know that dynamics can be represented on geometric object called phase space where both momenta and positions are taken into account. Dynamics is described by a trajectory on such a space. In addition, when energy is conserved, on a phase space the motion happens on a given geometrical object like a torus or a sphere. Therefore, there is a deep connection between Newton mechanics and geometry in a phase space. William Rowan Hamilton discovered this two centuries ago. Therefore, what does it make deterministic the Newton mechanics? This comes out from the commutativity of momenta and positions. Otherwise, we have quantum mechanics and the structure of the phase space changes dramatically. Quantum mechanics is the consequence of the noncommutative geometry of the phase space and one has to cope with operators and Hilbert spaces that are not properly geometric objects as our intuition learned from ordinary experience. One works with small volumes in the quantum phase space. The order of magnitude is given by $\hbar$. As opposite to the continuity in the classical Newtonian phase space.

As stated above, a Riemann manifold is characterized by a metric that determines the geodesic curves[5] on it. It is a differentiable manifold. This means that the derivative is well defined on functions defined on the manifold. We can make a correspondence between the points on a Riemann manifold and an algebra of functions defined on it that we can call maps. These functions can represent the manifold itself as the geometric concept of a point does. Stated in a simpler way, we are moving from a pictorial representation to its algebraic elements as for the phase space we move from the geometrical structure of the motion like an orbit to functions representing it like coordinates and momenta. When one moves from such functions to operators (or q-numbers *a la* Dirac as Heisenberg did) one gets a noncommutative Riemann manifold.

Connes, Chamseddine and Mukhanov proved that such a noncommutative manifold comes out made by two kinds of elementary volumes and it is from here that the deep connection with stochastic processes starts. This means that the relation between Fourier and Schrödinger equations, formally given by a Wick rotation, has a deep physical meaning. Mathematically, it entails the introduction of a new class of stochastic processes: the fractional powers of a Wiener process [10].

In summary, the relation between the Fourier and the Schrödinger equations, through a Wick rotation, is not merely a mathematical curiosity but rather has deep physical implications as the world is structured as a noncommutative Riemann manifold on which the particles, able to feel its quantization, perform Brownian motion with a process equation, which is exactly the Schrödinger equation. This means that the analogy between these famous equations relies on the kind of atoms one considers: For Fourier, it is the matter being parceled while for Schrödinger, it is the space being made by elementary volumes acting like the atoms in the matter. However, geometry is continuous in the former and discretized in the latter. Consequently, in a lattice world, if one moves randomly between the sites of the lattice, one obtains the solution to the Fourier equation. In a quantum world, represented - for instance - by a quantum Tartaglia-Pascal triangle [2], one has that the space on which the particle moves is discretized and so, moving on such a space yields the solution to the Schrödinger equation. The quantum Tartaglia triangle is, in a sense, the square root of the classical one [2] and this is the root of the "i" factor[4].

In the remaining part of the paper, we will discuss this deep connection showing how the emergence of a quantum world comes out from a quantized space.

II. THE WORLD IN PARCELS

As shown by Chamseddine in [11], we can always define a set of functions $Y^1, Y^2, ..., Y^N$ to cover a volume with the condition $(Y^1)^2 + (Y^2)^2 + \cdots + (Y^N)^2 = 1$. These represent spheres in $N + 1$ dimensions. We can cover the volume of a given manifold with such spheres so that they becomes a way to "measure" the manifold itself. However, unfortunately, we will find holes everywhere in our covering notwithstanding the manifold we started with was simply connected. Therefore, we cannot claim we are able to fully recover the original manifold from smaller volumes.

In order to overcome this difficulty, let us return to the case of quantum mechanics. As we stated above, in this case, we are studying the motion of a particle in a particular manifold represented through the coordinates $(q,p)$ with $q$ being the position and $p$ the momentum and this easily generalizes to the $N$ dimensional case. This is what physicists call *phase space*. William Rowan Hamilton introduced this concept earlier in the XIX century by reformulating Newton equations of motion in this way [12].

---

[5] A geodesic curve is a generalization of the concept of "straight line" of the Euclidean space to curved spaces as the shortest path joining two points.

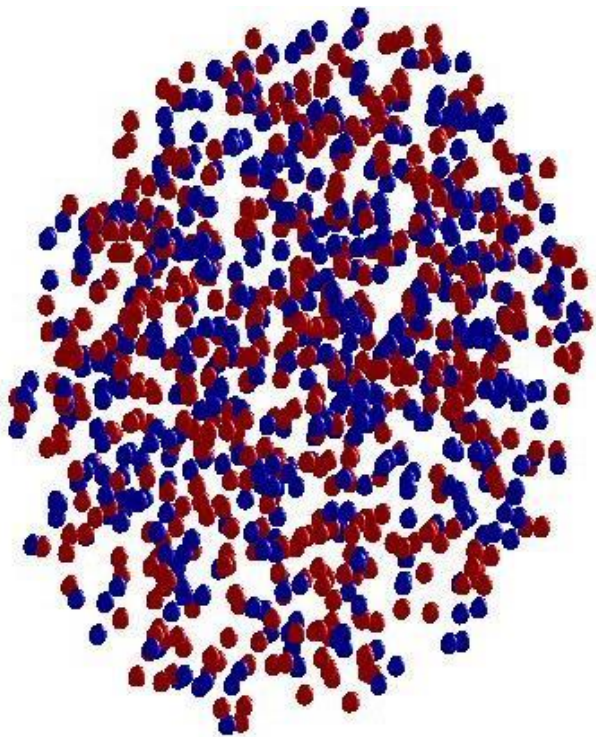

Fig. 1. A parceled manifold[6].

Furthermore, Heisenberg noticed that the coordinates in phase space do not commute at all and the reason of the non-commutation must be attributed to the existence of the Planck constant $\hbar$ [13]-[14]. This is generally stated in the form $[q,p] = i\hbar$. So, one can ask if there exists a solution to this equation in such a way to formulate the dynamics of a particle that moves on such noncommutative phase space. Today, we know the answer is affirmative, provided $q$ and $p$ are not ordinary c(omplex)-numbers but rather q(uantum)-numbers that is, self-adjoint operators that act on a Hilbert space of functions. So, we have a so-called "triple", in the nomenclature due to Alain Connes, formed by an algebra of operators, a Hilbert space of functions on which they act upon and we can postulate a spectral (Dirac) operator to measure distances in such a noncommutative phase space that acts on the functions in the Hilbert space. To define the Dirac operator one could use e.g. the Schrödinger equation that determines the way a particle moves around sensing distances. We see that we have built a geometry without recurring to any ordinary concept of point, geodesic and similar but our definition of a noncommutative phase space, a geometric concept anyway, is purely algebraic in a perfect Cartesian spirit.

---

[6] This figure was inspired by a talk given by Alain Connes in Castiglioncello (Italy) on 2014.

What are the consequences of having such a noncommutative structure for the phase space? So, let us consider a particle of unitary mass moving under the effect of an elastic spring. It is everyday experience that such a particle undergoes oscillator motion. This is true if the phase space is a standard geometric object. In such a case, the manifold on which the particle moves in phase space has the form of an ellipsoid, degenerating to an ellipse in two dimensions, geometrically characterized by the energy available to the particle. The effect of the Heisenberg commutation relation changes this dramatically as we get a fully quantized manifold in small volumes having a well-defined magnitude. This is standard material in quantum mechanics textbooks (e.g. see [15]). Specifically, if the available energy is $E$, we will have a number of quanta $n$ proportional to it and each quantum is measured by the product of $\hbar$ multiplied by the pulsation $\omega$ of the oscillation of the classical particle. Then, we can evaluate the volume of the quantized ellipsoid that in this simple case is just given by $\pi ab$, being $a = \sqrt{2E}$ and $b = \sqrt{\frac{2E}{\omega^2}}$ its semi-axes. Now, using energy quantization, the volume of the noncommutative manifold will be $2\pi(n + 1/2)\hbar$ and the phase space is quantized in small volumes each one of dimension proportional to the Planck constant.

Turning back to the Chamseddine's example, one can always try to cover a given manifold with small volumes using spheres but the covering will always be unsatisfactory. However, looking at quantum mechanics, we see that the Heisenberg quantization condition grants us that a perfect covering of a noncommutative manifold can be obtained. So, Chamseddine's example can be made at work moving to a noncommutative manifold. But, what should the quantization condition be in this case? To understand this, let us consider the case of a circle of radius 1. In this case, we can cover the circle with phase changes $Y = e^{i\theta}$ and we can choose an arbitrary angle to obtain the covering with even smaller angles. A distance could be given by $d(\theta,\theta') = |e^{i\theta} - e^{i\theta'}| = 2 \cdot \sin\left(\frac{\theta - \theta'}{2}\right)$. Now, if $D$ is the derivative with respect to the angle and promote the angle to a position operator, it is not difficult to observe that $Y^*[D,Y] = 1$ appears to be fully analogous to the Heisenberg quantization condition used above. The quantum problem informs us that the solution has a spectrum of integers for $D$. If we try to extend this to a generic one-dimensional Riemann manifold $M$, for $Y$ unitary as for the circle, the quantization condition $Y^*[D,Y] = 1$ will give a solution for $Y$ provided the length of the manifold satisfies $|M| \in 2\pi\mathbb{N}$, that is the Riemann manifold is quantized exactly as happened for the oscillator manifold. We have again a triple made by an operator algebra, a Hilbert space and a "Dirac" operator describing it. Now, we are covering a noncommutative manifold and the problem with an ordinary manifold applies no more.

These ideas can be generalized to arbitrary dimensions and what Connes, Chamseddine and Mukhanov proved was that the quantization condition that we have introduced with the example of the circle can be consistently applied in 4 dimensions. This permits the introduction of a fundamental experimental fact observed in nature: For each particle seen in nature, there is its anti-particle [16]. This fact is widely known as the matter-antimatter paradigm. However, for sure, the Universe would not exist the way we know and are able to observe without this double nature of the matter. To describe this, one has to introduce complex numbers. The effect of complex conjugation and time reversal moves the description from matter to anti-matter. This has a deep geometrical origin as shown in noncommutative geometry, as we need to consider a complex Riemann manifold to fit matter and anti-matter in place. This implies that a noncommutative Riemann manifold splits up in two kinds of a large number of parcels making up its volume as depicted in Fig. 1. We assume that these can be distributed in a random way because all the configurations are admissible for the building of the manifold yielding an identical result for the volume. So, we will be able to obtain a noncommutative Riemann manifold back provided we consider small parcels with volume 1 and $i$, the two fundamental unities in the world of mathematics.

### III. MOVING AROUND THE PARCELS

We can imagine that, for a particle with a given energy, moving on a manifold like that in Fig.1 means performing a kind of Brownian motion. Indeed, we can get a set of equivalent manifolds, each one with the same volumes but with the parcels differently disposed.

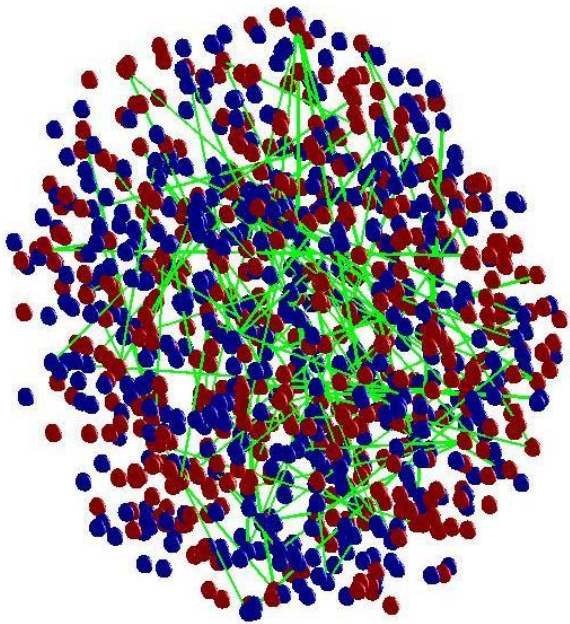

Fig. 2. Brownian motion on a parceled manifold.

To understand what is going on, we extend the case of the circle in the preceding section to the sphere. In this case we have to introduce two different operators $U_1$ and $U_2$ in full analogy with the case of the circle of the preceding section. We would like also to have a manifold that could be quantized consistently. This is generally possible only if one makes use of a Clifford algebra [16] . A Clifford algebra is a tricky way to take the square root of the wave equation. This was uncovered by Paul Dirac that in this way obtained the correct equation to describe the electron and a proper generalization of the Schrödinger equation that could account for the relativity discovered by Albert Einstein [17]. The Dirac equation is essential in noncommutative geometry to measure distances.

Consider the wave equation on the surface of the sphere of radius 1. This can be written as: $\partial_t^2\psi - \partial_\theta^2\psi - (\sin\theta)^{-2}\partial_\phi^2\psi = 0$. Taking the square root means that we have to use the nabla operator $\nabla$ and we do this by introducing three new objects $\sigma_1$, $\sigma_2$ and $\sigma_3$ to be defined below. Then, let us consider the new (Dirac) operator built as $D = \sigma_3\partial_t - i\vec{\sigma}\cdot\nabla$ where we have set $\vec{\sigma} = (\sigma_1, \sigma_2)$. One gets, taking the square $D^2$, that the wave equation is fully recovered provided σs have the properties $\sigma_1^2 = \sigma_2^2 = \sigma_3^2 = I$ and $\sigma_i\sigma_j + \sigma_i\sigma_j = 0$ when $i \neq j$. This is what we call a Clifford algebra and the lower non-trivial dimensionality for the σ are 2x2 matrices firstly introduced by Wolfgang Pauli to describe the spin of particles [18].

Given this Clifford algebra of σ matrices, we can build a quantized manifold by considering the map $Y = -i\gamma_1 U_1 - i\gamma_2 U_2 + \gamma_3 U_3$ obtained by analogy from the square root of the wave equation and the case of the circle in the preceding section. Here we have introduced a new Clifford algebra, the same as above, but independent of it as this applies now to the "coordinates" $U_1$, $U_2$ and $U_3$. This is the reason why we called these matrices γ rather than σ. Then, as Connes, Mukhanov and Chamseddine proved, a manifold gets quantized with respect to the conjugate operator $D$ provided the Heisenberg-like quantization condition $Tr(Y[D,Y][D,Y][D,Y])=\sqrt{\kappa}$ σ holds, being $\kappa = \pm 1$ and σ a 2x2 matrix [7]-[8]. The trace is over the γ matrices. This quantization condition, which we have seen to arise naturally from the case of the circle of the preceding section, represents a generalization of the well-known Heisenberg condition $[q,p] = i\hbar$ we discussed in the introduction for quantum mechanics and the phase space. The triple product of $[D,Y]$ is due to the dimensionality of the manifold we are now considering. What Connes, Mukhanov and Chamseddine show is that this quantization condition admits a solution, i.e. $U_1$, $U_2$ and $U_3$ exist, if and only if the volume is quantized. This is true given two kinds of small volumes (parcels): One has volume 1 and the other volume $i$.

The question is how will a particle move on such a manifold? What will its equation of motion be? The answer was given in a recent paper by one of the authors [19]. To understand this, it is important to notice that this noncommutative manifold will have randomly distributed parcels that make it as, whatever configuration we will choose, the volume of the manifold will remain the same. This situation is really similar to the Brownian motion of a particle in a fluid whose theory was put forward by Albert Einstein in 1905 [4]. He showed that the motion of a particle, under the effect of the scattering of the molecules composing the fluid, is ruled by the Fourier heat equation. This means that this is a stochastic process with a given probability distribution function. In our case, there is something different as the motion of the particle, although random as well, can hit both real and complex valued volumes randomly. This implies that our stochastic process cannot be a real one but, rather, a complex valued stochastic process.

In order to have a stochastic process producing the values 1 and $i$, we start with a Bernoulli process $B$ that describes the tossing of a coin, yielding a random sequence with +1 and -1. From this we can introduce the process $\Phi = \frac{1+B}{2} + i\frac{1-B}{2}$ that will yield a similar sequence with 1 and $i$. It is not difficult to observe that $\Phi^2 = B$. Then, the particle will move around the maps $Y$ that make the manifold and these are members of a Clifford algebra as we have seen. Therefore, this is a Wiener process but also means that the Bernoulli process $B$ is not independent but will originate from the signs of the single Brownian steps. Using the typical notation of stochastic calculus, where steps are written like differentials, the case of our sphere will take a form like (see [19]) $dY = [\gamma_2(k + a \cdot dt - ib \cdot dU_2 B_2)\Phi_2 + \gamma_1(m + c \cdot dt - ie \cdot dU_1 B_1)\Phi_1 + i\gamma_0(f\Phi_1 + g\Phi_2)]$ where the condition $Y^2 = \pm 1$ must be understood and $a, b, c, e, f, g$ and $k, m$ are numbers. We used indexes on the various Bernoulli processes to distinguish their contributions on the different maps as shown in [10] and [19]; we are just extracting the square root of a Wiener process. This entails the use of a Clifford algebra (the Dirac's trick) as happened for the wave equation.

So, summing everything up, we have seen that a Riemann manifold can be quantized if it is noncommutative and, to accommodate the matter-antimatter paradigm experimentally, this quantization has parcels of unitary volumes 1 and $i$. Moving on it entails a complex stochastic process that is nothing else than the square root of the well-known Brownian motion (see Fig. 2). Now, we should ask what is the equivalent of the Fourier heat equation for this case as this is a characteristic of whatever stochastic process. The surprising answer is that now, for a complex stochastic process, we have a complex Fourier equation. The magic happened and we are back to Schrödinger.

IV. THE SURPRISING RELATION BETWEEN THE SCHRÖDINGER AND FOURIER EQUATIONS

From the discussion given above it appears that the Schrödinger equation is a Fourier equation in disguise. They both represent a stochastic processes but one has to move from real stochastic processes to complex ones. Anyhow, the relation is deeper as one arises taking a formal square root of the other. In the end, one has a diffusion equation for the random process but, in a quantum world, the equivalent of the probability distribution function is not real and one has to struggle working with modulus square of what is commonly called a wave function that is now the probability distribution function to consider for natural processes.

This can be put at test very easily with a numerical computation. We compute a Brownian motion and take its square root. The former has a histogram recovering the Fourier kernel, a Gaussian distribution. We can Wick rotate the numerical data of the square root yielding a Gaussian curve again, the Wick-rotated Schrödinger kernel, as expected by our connection [19]. The result is given in Fig. 3 and is in excellent agreement with expectations.

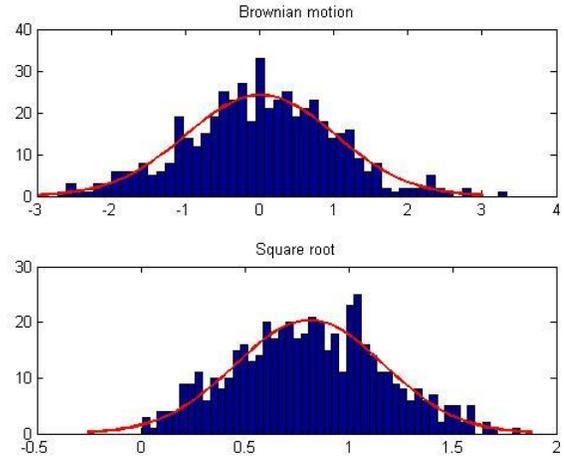

Fig.3. Comparison between the Fourier kernel of a Brownian motion and the Wick-rotated Schrödinger kernel of its square root.

So, the formal change of the time variable $t \to it$, introduced by Wick, hides a deep physical fact: The ordinary Brownian motion described by the Fourier equation changes into the motion on a parceled universe, where matter and antimatter exist, particles spin and the Schrödinger equation describes what is going on. But now, we are working with complex quantities whose only the modulus square can make sense. Fig. 4 depicts this relation between Fourier and Schrödinger equations and related physical consequences.

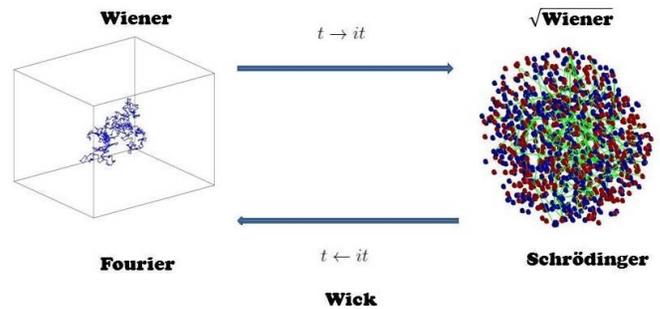

Fig. 4. How Fourier and Schrödinger speak each other and Wick invented a communication way.

V. CONCLUSIONS

We have seen how a simple formal relation between two famous equations entails a deep mathematical concept with new stochastic processes underlying it. A new mathematical technique also forecasts a wealth of possible applications.

Indeed, this paper may benefit modern applications of quantum mechanics to engineering in communication and sensing. Namely, quantum communications through entangled states [20], that are solutions of the Schrödinger equation we derived, from a new class of complex stochastic processes, and quantum radars [21] that can benefit from quantum illumination [22]-[23]. Entangled states are particular solution of the Schrödinger equation with two or more particles like photons. Representing them through stochastic processes could make easier their use in designing devices that employ such states.

Stochastic processes like the ones that are discussed here can be easily simulated digitally or through some analogic device using discrete components. In this view, filtering can witness new avenues of applications. In general, there could be a lot of possible new applications wherever a new class of stochastic processes is uncovered. This, in view, is our hope for the future for this exciting mathematical achievement we have got from a world so distant from our common sense.

**Marco Frasca** is currently employed at MBDA Italia S.p.A. working on signal processing and sensing. He is also a theoretical physicist with more than 80 publications on refereed journals.

**Alfonso Farina** in 1974, he joined Selenia, then Selex ES, where he became Director of the Analysis of Integrated Systems Unit and subsequently Director of Engineering of the Large Business Systems Division. In 2012, he was Senior VP and Chief Technology Officer of the Company, reporting directly to the President. He retired in October 2014. From 1979 to 1985, he was also professor of "Radar Techniques" at the University of Naples (IT). Today, He is a Visiting Professor at University College London (UCL), Dept. Electronic and Electrical Engineering. He is a Distinguished Lecturer of IEEE AESS and IEEE SPS Distinguished Industry Speaker. He is a consultant to Leonardo S.p.A. "Land & Naval Defence Electronics Division" (Rome). He is a Member of the Editorial Board of IEEE SP Magazine.